\documentclass[11pt,twoside]{article}
\usepackage{asp2010}
\usepackage{graphicx}

\resetcounters

\begin{document}

\title{Magnetic Fields and the Formation of Cores and Disks}
\author{Shantanu~Basu,$^1$ Nicole~D.~Bailey,$^1$ and Wolf~B.~Dapp$^2$}
\affil{$^1$Department of Physics and Astronomy, Western University, London, Ontario N6A 3K7, Canada}
\affil{$^2$J\"ulich Supercomputing Centre, Institute for Advanced Simulation, FZ J\"ulich, Germany}

\begin{abstract}
We review recent results of non-ideal magnetohydrodynamic models for the 
fragmentation of molecular clouds and the collapse of cloud cores to 
form protostar-disk systems. Thin-disk models can elucidate many aspects of
the physical problem and allow the calculation of large dynamic range of 
time and length scales. 
\end{abstract}

\section{Introduction}
A large-scale magnetic field exerts a dramatic influence on fragmentation 
properties of molecular clouds, especially when accounting for non-ideal
magnetohydrodynamics (MHD) based on partial 
ionization effects. Furthermore, within a collapsing cloud core, the 
magnetic field assumes a crucial role in the near-protostar environment, 
a region where the magnetic flux problem is resolved and the angular 
momentum content of the forming star-disk system is finally determined. 

The magnetic field is most generally invoked in star formation as an explanation for the observed very low efficiency of star formation in molecular clouds.
Molecular clouds are the exclusive sites of star formation but are characterized by a very low efficiency of star formation, e.g., in the Taurus Molecular Cloud only $\sim 1$\% of the mass is in the form of stars \citep{gol08}, and the global rate of star formation in the Galaxy is estimated by various means to be 
$1 - 5\, M_{\odot}$ yr$^{-1}$ \citep{mis06,rob10}. This is only $\sim 1\%$ of that expected if molecular gas in the Galaxy is collapsing on its dynamical time.

Magnetic fields that create a subcritical mass-to-flux ratio are a 
viable means of
preventing most molecular cloud material from forming stars during the lifetime of the cloud, assuming that the neutral-ion coupling is sufficiently strong.
In this paper, we will frequently refer to the normalized mass-to-flux ratio
$\mu \equiv 2 \pi G^{1/2} \sigma_n/B_z$, where $(2 \pi G^{1/2})^{-1}$ is the
critical value of the mass-to-flux ratio. Clouds that are subcritical 
($\mu < 1)$ cannot collapse in the flux-freezing limit but more
realistically undergo an ambipolar-diffusion-initiated fragmentation 
that occurs on a time scale many times longer than the dynamical time. 
Clouds that are supercritical ($\mu > 1$) will fragment rapidly on a
dynamical time into Jeans mass scales. The transcritical case 
($\mu \approx 1$) is a fascinating middle ground that is discussed
in this paper. 

We are working in the scenario that the low efficiency of star formation
is explained by having most of the diffuse envelope of a molecular cloud
in the form of subcritical gas. This subcritical common envelope is not
expected to lead to significant (or any) star formation during the lifetime
of the molecular cloud. However, regions of the molecular cloud may become
transcritical or supercritical due to accumulation of matter into the 
molecular cloud or nonlinear perturbations that lead to localized
rapid ambipolar diffusion. These can allow fragmentation
to be initiated in subregions of a molecular cloud, and lead to weak or
rich clusters of stars.

Once cores have formed by a fragmentation process, they are mildly 
supercritical and begin runaway collapse toward formation of a central
protostar. The angular momentum provides a barrier to this, with 
observationally determined ratios of rotational to gravitational
energy $\beta \approx 10^{-4}$ still leading to the formation of large disks
if angular momentum is conserved. However, even a 
mildly supercritical mass-to-flux ratio provides enough magnetic coupling
to the core envelope that disk formation is suppressed in the flux-freezing 
limit. This is known as the {\it magnetic braking catastrophe}.
We show how this problem is resolved using a detailed treatment of
microphysics including a chemical network in order to calculate
coefficients of Ohmic dissipation and ambipolar diffusion.
Exotic explanations are not required for a small disk to form at
early times.

\section{Fragmentation of Magnetic Molecular Clouds}

Numerical simulations in the thin-disk approximation have revealed the
dramatic dependence of fragmentation properties on the ambient initial mass-to-flux
ratio $\mu_0$ \citep{bas09b}. Unlike a uniform medium, sheets and filaments 
exhibit a preferred
scale of fragmentation for any combination of mass (per unit area or length).
When a magnetic field and partial ionization are
taken into account, the regions with transcritical mass-to-flux ratio have
the largest preferred fragmentation scales. This is clearly seen from the
results of a linear perturbation analysis of a partially ionized sheet. 
Figure 1 shows the run of preferred fragmentation scale versus mass-to-flux
ratio, taken from \citet{bai12} and based on the earlier results of 
\citet{cio06}. The peak actually occurs at $\mu_0 = 1.1$,
slightly on the supercritical side, as these modes are driven by a combination
of neutral-ion slip and inward dragging of field lines. The latter allows 
strong restoring forces to be set up due to the tension and pressure of the
relatively strong magnetic field. The solid line in Figure~1 is the set of
preferred wavelengths in the limit of flux freezing (neutral-ion collision
time $\tau_{\rm ni,0}=0$) and the other lines are the set of preferred
wavelengths if there is partial coupling of neutrals and ions with successively
decreasing ionization fraction as $\tau_{\rm ni,0}$ increases. The case
$\tau_{\rm ni,0}/t_0=0.2$ corresponds to the canonical cosmic ray 
ionization rate.

\begin{figure}
\plottwo{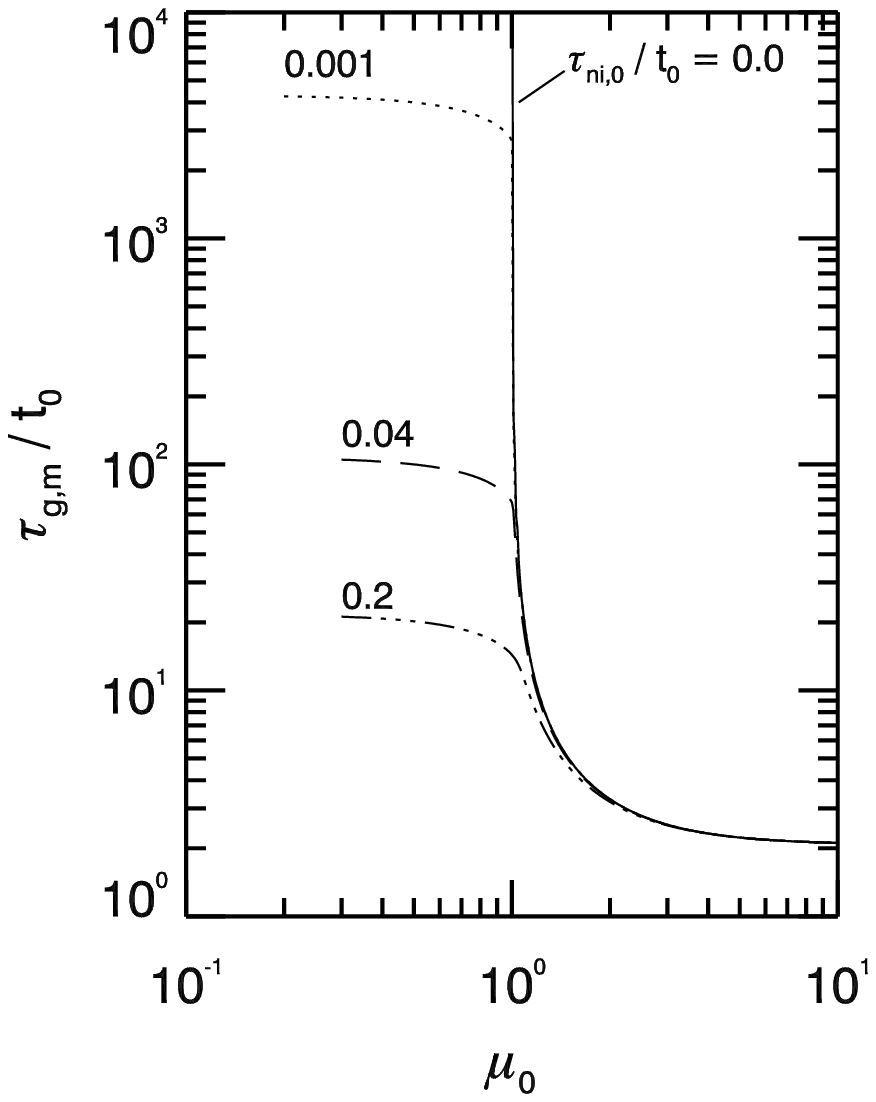}{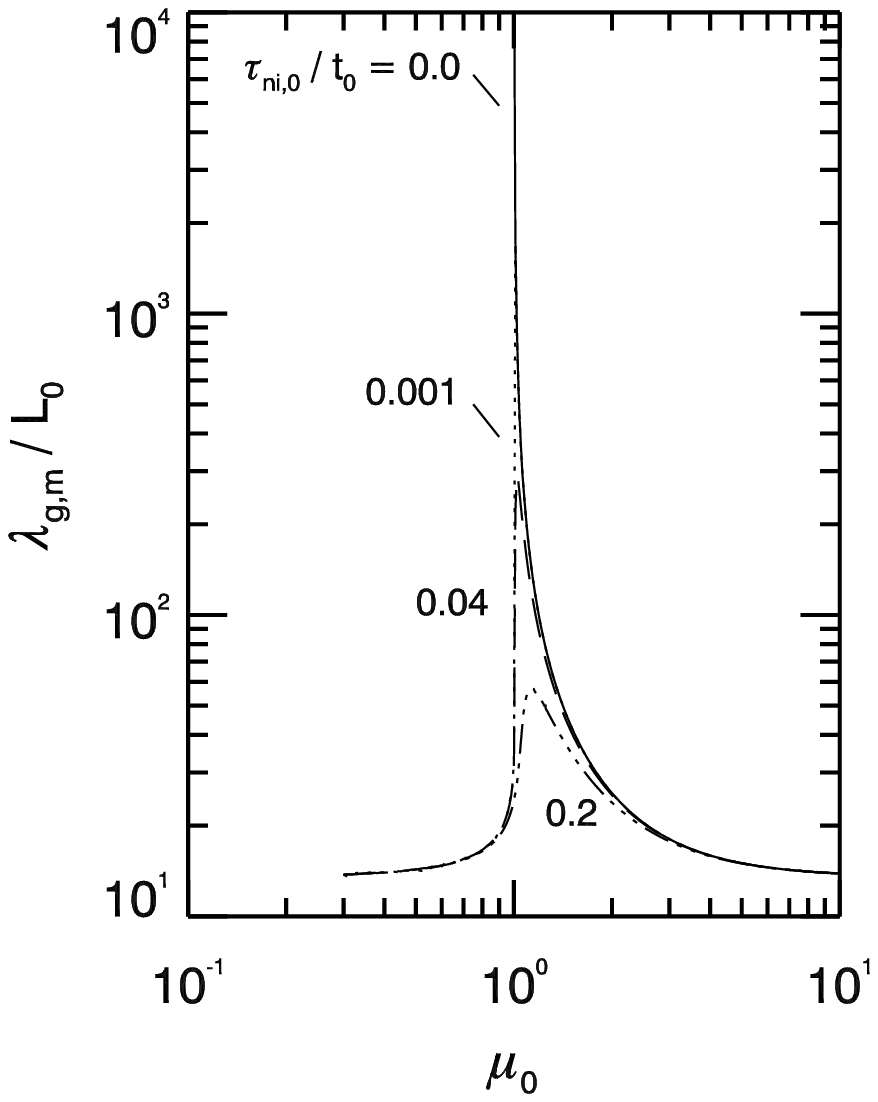}
\caption{
Left: Minimum growth time of the most gravitationally unstable mode ($\tau_{\rm g,m}/t_0$) as a function of mass-to-flux ratio ($\mu_0$). Right: Length
scale of the most unstable mode ($\lambda_{\rm g,m}/L_0$) as a function of $\mu_0$. Units of time $t_0$ and length $L_0$ are described in \citet{bas09b} and are related to the dynamical time and Jeans scale for the system.
Different values of $\tau_{\rm ni,0}$ correspond to different levels of ionization, with $\tau_{\rm ni,0}/t_0=0.2$ corresponding to the canonical cosmic ray
ionization rate, and $\tau_{\rm ni,0}/t_0=0$ corresponding to the flux-frozen limit. From \citet{bai12}.
}
\end{figure}

Figures 2 and 3 show images of column density along with velocity vectors for
fragmentation of clouds that are significantly subcritical ($\mu_0 = 0.5$),
transcritical ($\mu_0=1.1$), somewhat supercritical ($\mu_0=2.0$), and
significantly supercritical ($\mu_0=10$). All modeled regions are the
same size, about 2 pc $\times$ 2 pc for typical parameters. The transcritical
cloud yields the largest fragments, while the highly sub- and supercritical
clouds yield the smallest ones. The latter weak magnetic field case yields
the most elongated structures. 

An extrapolation of these numerical results is that a cloud with a variety
of mass-to-flux ratios in its various regions will develop a broad
core mass function. Monte Carlo simulations of this using the linear
fragmentation theory \citep{bai13} reveal that a narrow lognormal distribution of
cores in the hydrodynamic limit is transformed to a broad distribution
with a shallow power-law high-mass tail in the magnetic flux-frozen case,
and that the addition of ambipolar diffusion leads to a broad mass function
with a high-mass cutoff. 

\begin{figure}
\plottwo{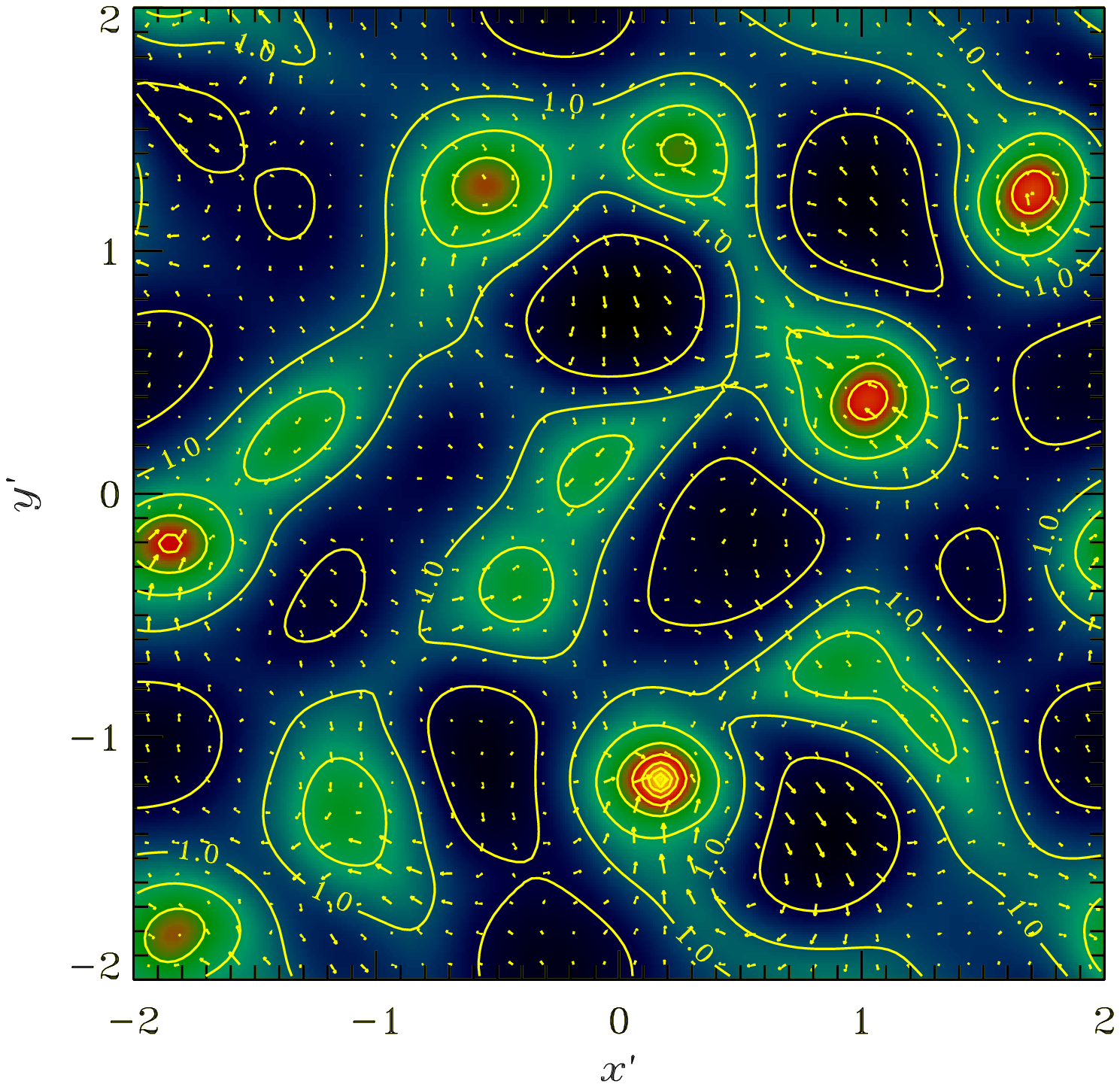}{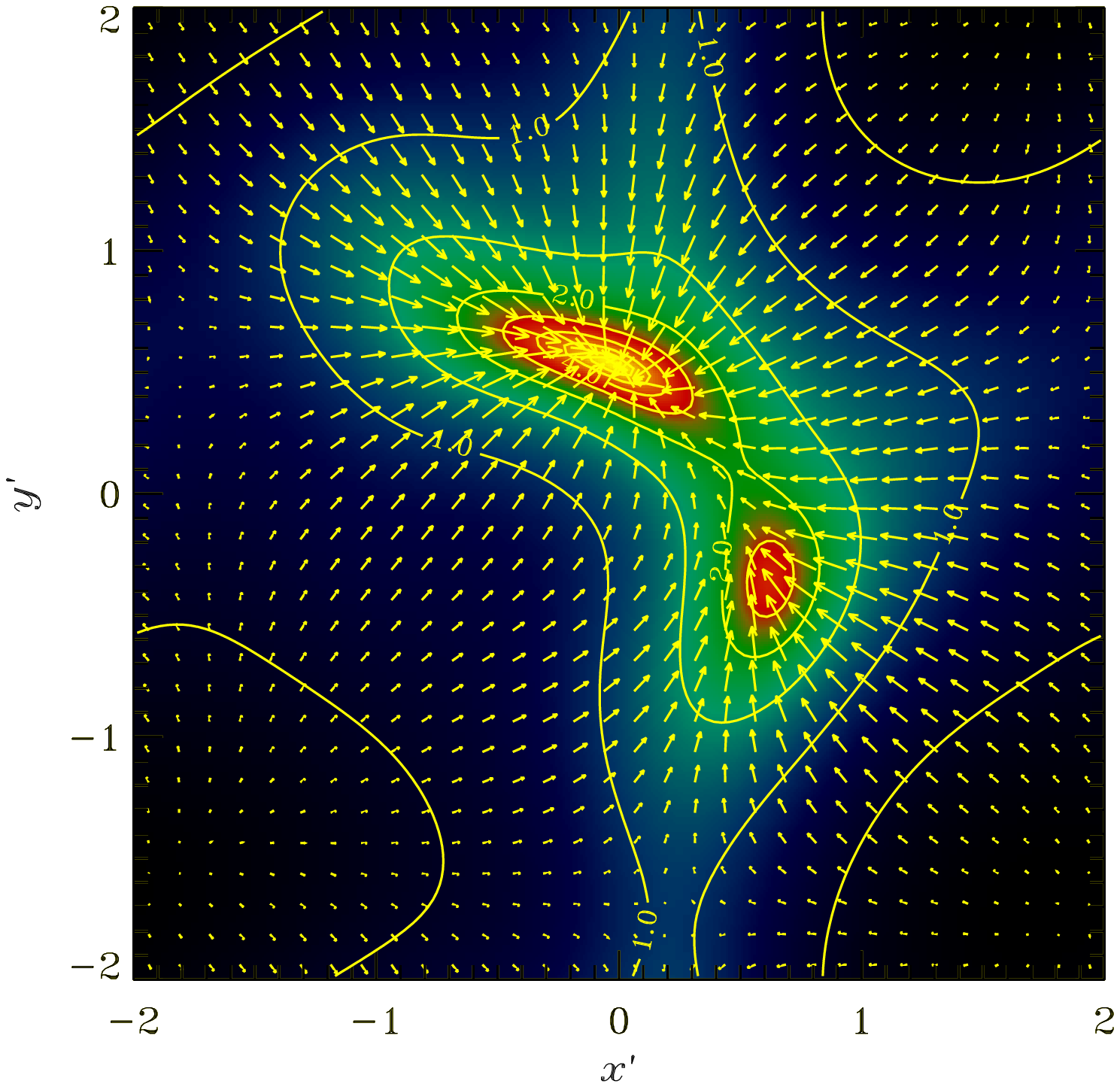}
\caption{Image and contours of column density $\sigma_n(x,y)$ and velocity
vectors of neutrals, at the time when the column density is enhanced by a factor of 10 and 
runaway collapse is well underway. Left: subcritical (magnetically dominated) model with
$\mu_0 = 0.5$. Right: transcritical model with $\mu_0 = 1.1$. The horizontal or 
vertical distance between footpoints of velocity vectors corresponds to a 
speed of 0.5$c_s$, where $c_s$ is the isothermal sound speed. From \citet{bas09b}.}
\end{figure}

\begin{figure}
\plottwo{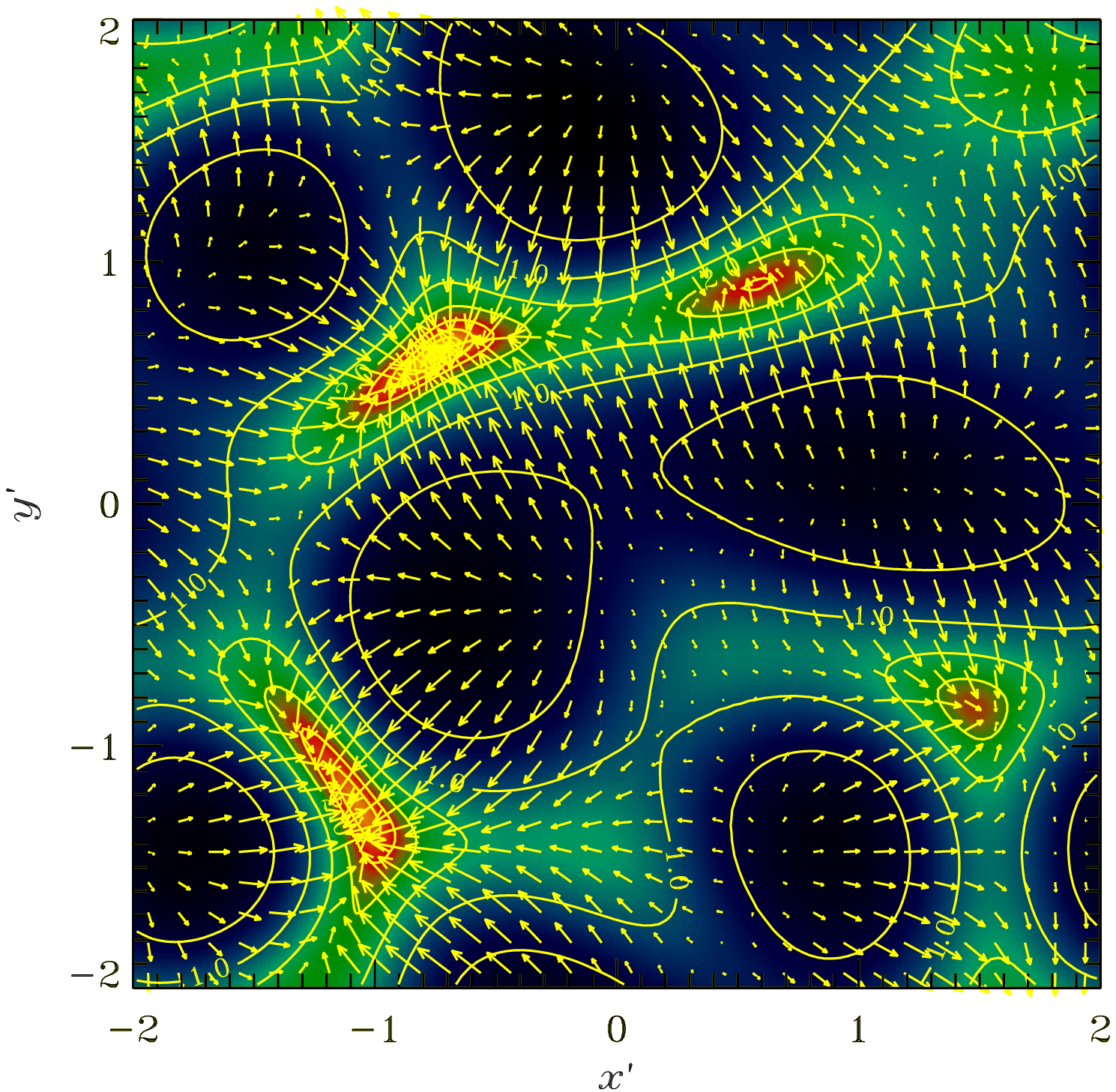}{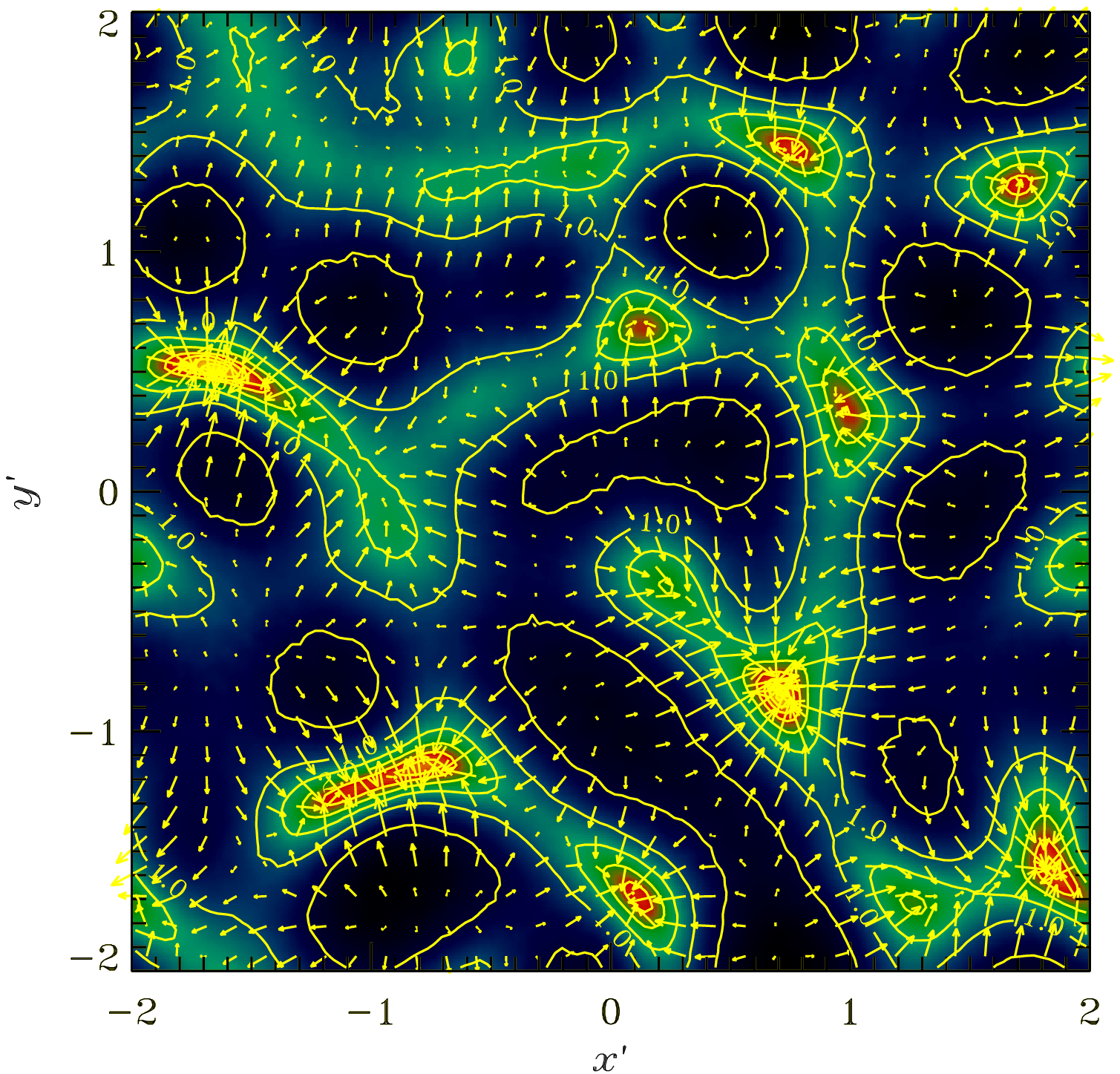}
\caption{Same as previous figure but for different models. Left: Mildly supercritical
model with $\mu_0 = 2.0$. Right: Highly supercritical model with $\mu_0 = 10$. From \citet{bas09b}.}
\end{figure}

While the thin-disk approximation is an idealization of a molecular cloud,
it is motivated by the idea that while molecular clouds will in general
have rarefied turbulent envelopes, the star formation will take place
in embedded dense sheets and filaments with near-thermal line widths. 
This is illustrated
theoretically by the one-dimensional simulations of turbulence propagation
in a stratified atmosphere by \citet{kud03,kud06}. They show that large nonlinear
motions occur in the rarefied regions but that a dense midplane region maintains
transonic motions.
Furthermore, three-dimensional simulations of flattened clouds without
vertically propagating turbulence
\citep{kud07,kud08,kud11} are also consistent with the results of the
thin-disk simulations. We do not focus on the three-dimensional 
results in this paper.

The preferred fragmentation scale will also depend on the level of ionization,
and in fact a dramatic drop in this quantity is expected within a molecular
cloud in the region where the column density is high enough to shield out
the background ultraviolet starlight. This leads to the idea of two-stage 
fragmentation \citep{bai12}, described as follows. One can envision two
dramatic events during the assemblage of a molecular
cloud. If we assume that a molecular cloud is assembled
from HI cloud material that is generally subcritical
\citep{hei05} and ionized by ultraviolet (UV) starlight,
then a cloud may resist fragmentation (due to the very
long ambipolar diffusion time) until flows primarily along
the magnetic field lines raise the mass-to-flux ratio to
slightly above the critical value. At this point, the transcritical
(but still UV ionized) cloud may form large fragments
as predicted by the linear theory. As these fragments
develop, they will also become more supercritical,
since their evolution is partially driven by neutral-ion
drift. Once the column density in the fragments also
crosses the column density threshold for the transition
to cosmic ray dominated ionization, the fragmentation
length and time scales will drop dramatically, and a second
stage of fragmentation may be possible. 

\begin{figure}
\includegraphics[scale=.50,angle=-90]{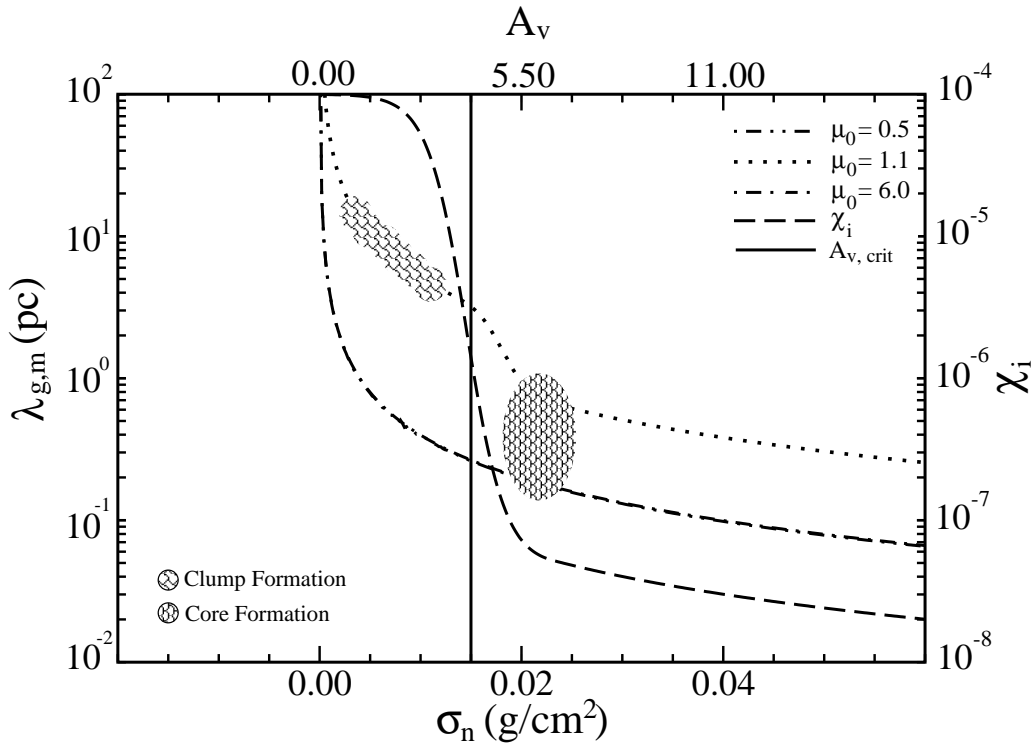}
\caption{Illustrative plot of the two-stage fragmentation scenario.
The ionization fraction $\chi_i$ versus column density is in the dashed line.
The other lines show the dependence of preferred fragmentation scale
on the column density for various values of $\mu_0$.
Transcritical clumps that form on the left side of the critical 
visual extinction $A_{\rm v,crit}$ (vertical solid line) for ultraviolet 
starlight can 
undergo a second fragmentation event on the right side of 
$A_{\rm v,crit}$ when the preferred fragmentation scale drops significantly.
Shaded regions show the parameter space for the formation of clumps and
cores respectively. From \citet{bai12}.
}
\end{figure}

Figure 4 shows a schematic picture of the two-stage
fragmentation model. It plots the ionization fraction $\chi_i$ as well as
the fragmentation length scale $\lambda_{\rm g,m}$ for various values
of $\mu_0$, all as functions of column density.
In the diffuse regions where clouds are just forming from subcritical gas,
we expect the mass-to-flux ratio be in the transcritical ($\mu_0 \approx 1$)
regime. The parameter space where we expect
the fragmentation of the gas into parsec size clumps is
indicated by the hatched region to the left of the critical 
visual extinction $A_{\rm v,crit}$.
Linear analysis reveals that the hatched region corresponds
to fragmentation time scales on the order of
2-10 Myr. As the column density of the region increases, 
the region will cross over to the
right hand side of $A_{\rm v,crit}$. The hatched region on this
side of $A_{\rm v,crit}$ represents the parameter space for subfragmentation
within the clump. The length scales of fragmentation here are a few $\times$
0.1 pc, roughly the size of observed dense cores. The time scales
of fragmentation are also significantly shorter than the 
timescale of collapse for the larger parent transcritical clump.
This second-stage fragmentation is expected to take place in gas that is
either still transcritical or has become supercritical.

\section{Core Collapse to Disk Formation}

\begin{figure}
\plottwo{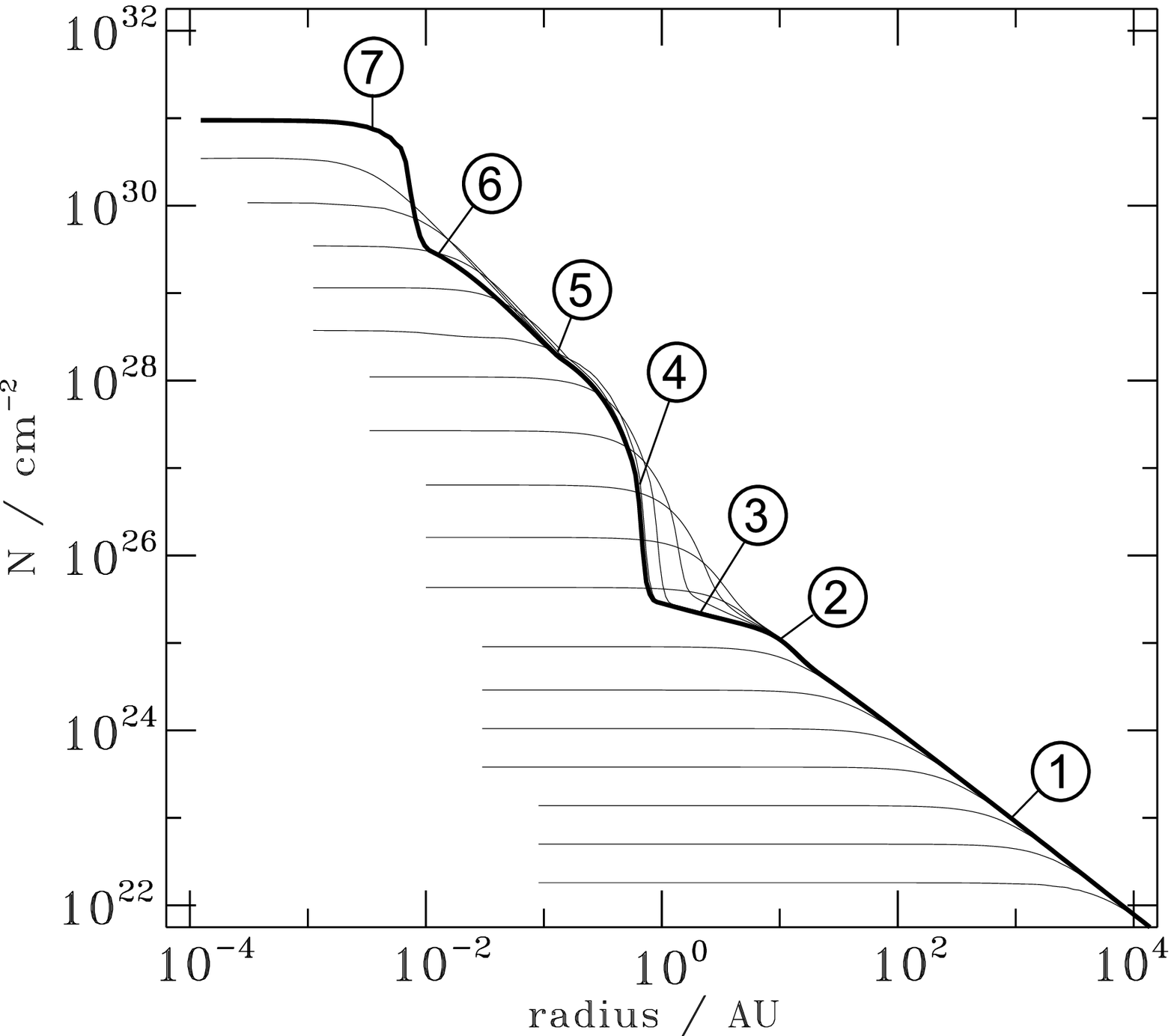}{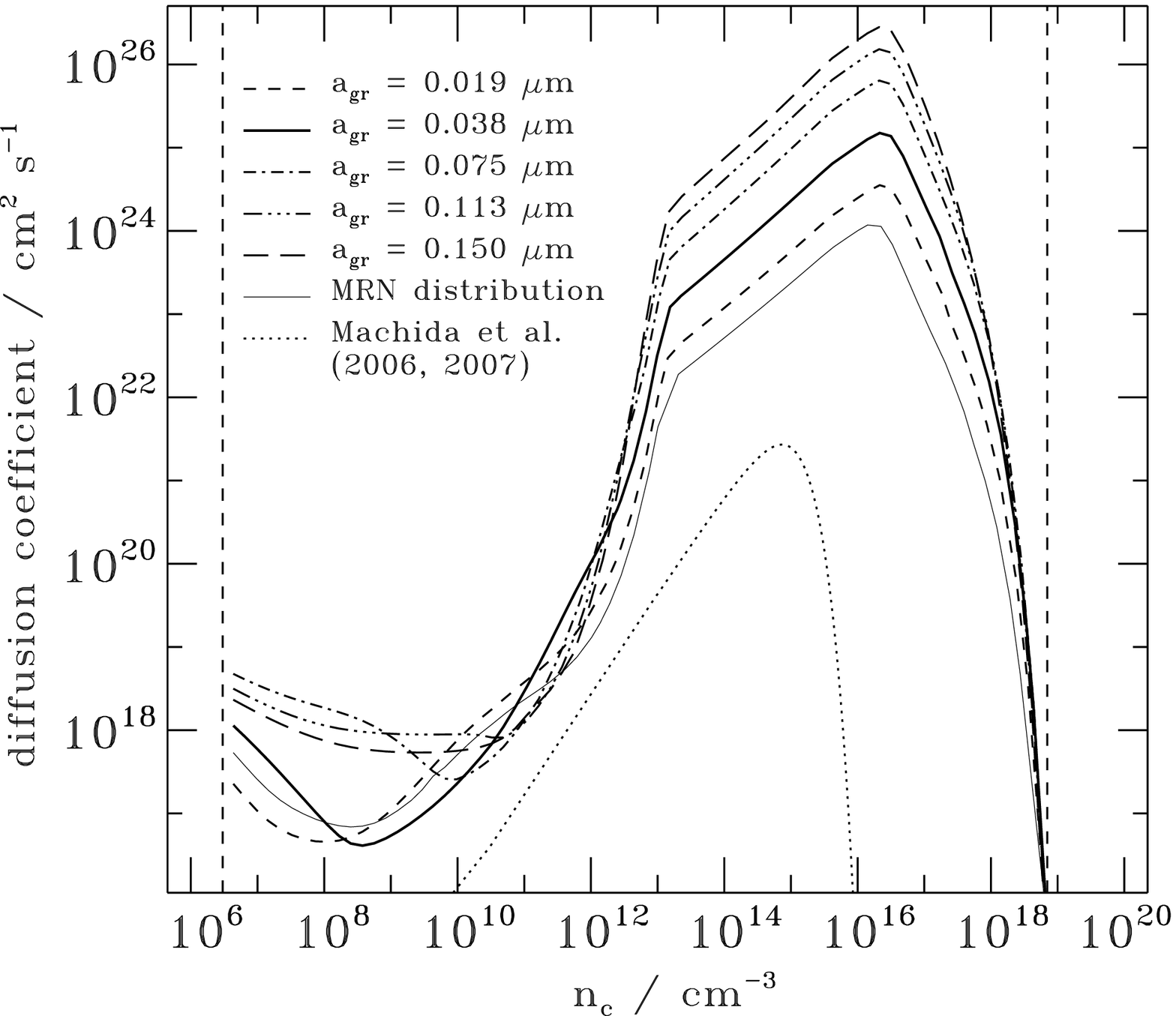}
\caption{Left: 
Column number density profile versus radius for collapsing core model. The thin lines (in ascending order) are plots at successive times.
Several features are identifiable via their associated breaks in the profile.
(1) Prestellar infall profile with $N \propto r^{-1}$. (2) Magnetic wall at $\approx 10~\mathrm{AU}$, where the bunched-up field lines decelerate material before it continues
the infall. (3) Expansion wave profile with $N \propto r^{-1/2}$ outside the first
core. (4) First core at $1~\mathrm{AU}$. (5) Infall profile onto the second core with $N \propto r^{-1}$. After the first core has reached $\approx 1,000~\mathrm{K}$, it starts to collapse, as H$_{2}$ is dissociated. (6) Expansion wave profile with $N \propto r^{-1/2}$ outside the second core. (7) Second core at $\approx 1~R_{\odot}$.
Right: 
Central magnetic diffusion coefficient $\eta_{\rm eff}$ versus number density for
different grain sizes, extracted from the dynamical model. The vertical line at the left indicates the density at which the detailed chemistry and non-ideal MHD treatment is switched on. Beyond $n_{\mathrm{c}}\approx 10^{18}~\mathrm{cm}^{-3}$ the resistivity plummets, after having already declined due to thermal ionization. This is where we switch the chemistry calculations off again, and is denoted by the vertical line on the right. Due to grain destruction, flux-freezing is restored there.
Both figures from \citet{dap12}.}
\end{figure}

In order to form stars of the observed sizes and rotation rates,
most of the angular momentum has to be removed from the
infalling gas. This is the classical angular momentum problem.
The required reduction of angular momentum
is often credited to magnetic braking, which acts during the
contraction and collapse by linking the core with its envelope
and transferring angular momentum. Recent numerical simulations of
protostellar collapse under the assumption of ideal MHD 
have suggested the converse problem,
namely that cores may experience catastrophic magnetic braking.
This is the result that an extremely pinched magnetic field
configuration has enough strength and exerts a long enough lever
arm on the inner regions of collapse that magnetic braking can
suppress the formation of a centrifugal disk entirely.

Catastrophic magnetic braking was first demonstrated by
\citet{all03}, who used two-dimensional (2D) axisymmetric
calculations. Subsequent ideal MHD simulations by \citet{mel08}
in 2D and \citet{hen08} in three
dimensions showed that catastrophic magnetic braking
occurs for initially aligned (magnetic field parallel to rotation
axis) rotators in which the magnetic field is strong enough that
$\mu \leq 10$.

The left panel of Figure 5 shows the column number density 
profile versus radius at various times in the collapse process
of a prestellar core \citep{dap12}. Using the thin-disk approximation,
the collapse is followed over a wide dynamic range of scales. 
A detailed chemical network model is used
to calculate ionization levels and the effect of ambipolar diffusion
and Ohmic dissipation. Many important features of the collapse 
profile are captured in this model, as described in the figure caption.
The high resolution of the model allows to identify all important 
breaks in the profile including the presence of a double expansion wave and 
a magnetic wall.
The right panel of Figure 5 shows the effective magnetic diffusion 
coefficient in the central cell, for various assumed grain sizes
$a_{\rm cr}$ in the chemical network model, or for an MRN grain
size distribution. The diffusivity rises steeply starting at a
number density $n \approx 10^{12}$ cm$^{-3}$, leading to rapid magnetic
diffusion and a dramatic reduction of magnetic braking.
Three-dimensional simulations \citep{mac06,mac07} also show
very effective Ohmic dissipation in this phase of evolution.

\begin{figure}
\plottwo{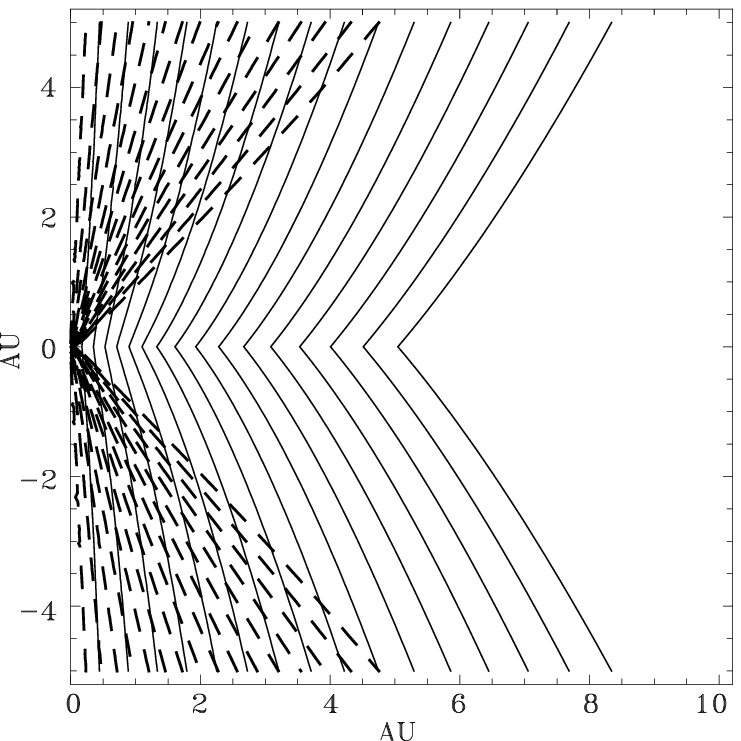}{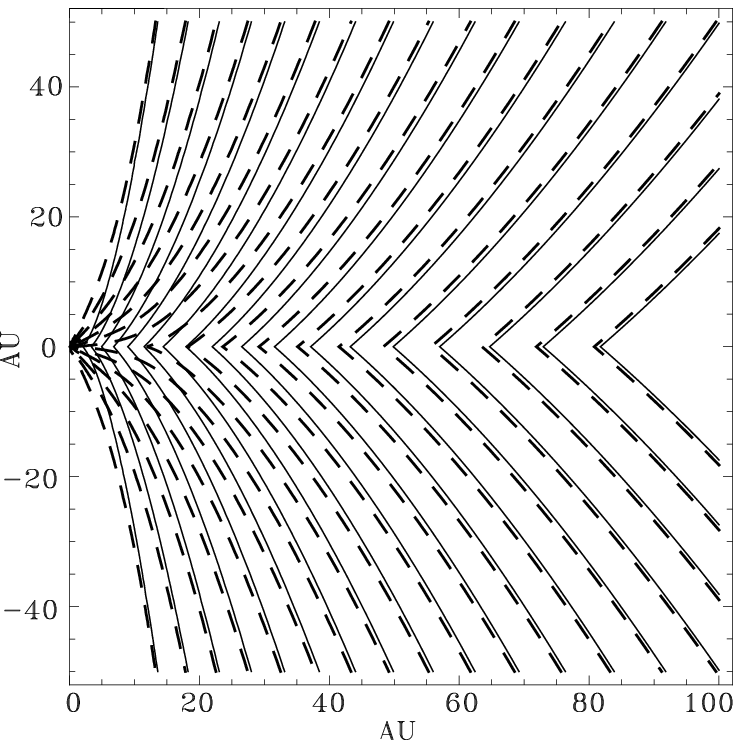}
\caption{
Magnetic field lines in the vicinity of a newly formed stellar core. The box on the left has dimensions $10~\mathrm{AU}$ on each side,
while the box on the right has dimensions $100~\mathrm{AU}$. Both figures from \citet{dap12}.}
\end{figure}

Figure 6 shows the magnetic field lines in the model, in boxes of size
10 AU and 100 AU, respectively. 
The dashed lines represent the flux-freezing model, while the solid lines show \textit{the same} field lines for the model including non-ideal MHD effects for a grain size $a_{\mathrm{gr}}=0.038~\mathrm{\mu m}$. In 
both cases, the second core has just formed and is on the left axis midplane. The field lines straighten out significantly on small scales in the non-ideal 
MHD model compared to the flux-frozen model.

Due to the magnetic diffusion and relaxed field line structure within 10 AU,
magnetic braking becomes ineffective and a protostellar disk is indeed 
able to form very soon after the formation
of the second core. The centrifugal support rises rapidly and a low-mass 
disk of radius $\approx 10\, R_{\odot}$ is formed at the earliest stage 
of star formation, when the second core has mass $\sim 10^{-3}\,M_{\odot}$. 
The mass-to-flux ratio is $\sim 10^4$ times the critical value in the 
central region.

Estimates based on the angular momentum in the collapsing core predict
that no disk larger than $\sim 10$ AU form around Class 0 objects
younger than $\sim 4 \times 10^4$ yr. This agrees well with the 
observations of \citet{mau10} who do not find evidence for
disks $\gtrsim 50$ AU in Class 0 objects. ALMA will soon allow 
observers to test this prediction more stringently.

\acknowledgements 
We thank the organizers of the ALMA symposium for a very stimulating conference
in the beautiful resort location of Hakone.

\bibliographystyle{asp2010}
\bibliography{basurefs}

\end{document}